\documentclass[manuscript]{aastex}

\slugcomment{Submitted to PASP May 10, 2002}

\shorttitle{ Rayleigh Laser Guide Star Systems:  UnISIS Bow Tie Shutter and CCD39
Wavefront Camera}

\shortauthors{Thompson, Teare, Crawford and Leach}

\begin{document}

\title{Rayleigh Laser Guide Star Systems:  UnISIS Bow Tie Shutter and CCD39 
Wavefront Camera}
\author{Laird A. Thompson\altaffilmark{1}, Scott W. Teare\altaffilmark{2,3}, 
Samuel L. Crawford\altaffilmark{4}, and Robert W. Leach\altaffilmark{3,5} }

\altaffiltext{1}{Astronomy Department, University of Illinois Urbana-Champaign, 1002 
W. Green St., Urbana, IL  61801; thompson@astro.uiuc.edu}
\altaffiltext{2}{Departments of Electrical Engineering and Physics, New Mexico Tech, 
801 Leroy Place, Socorro, NM  87801; teare@ee.nmt.edu}
\altaffiltext{3}{Astronomy Department, San Diego State University, San Diego, CA  
92182}
\altaffiltext{4}{Jet Propulsion Laboratory, California Institute of Technology, 
4800 Oak Grove Drive, Pasadena, CA  91109; slcrawford@huey.jpl.nasa.gov}
\altaffiltext{5}{Astronomical Research Cameras, Incorporated, San Diego State 
University, San Diego, CA  92182; leach@astro-cam.com}

\begin{abstract}Laser guide star systems based on Rayleigh scattering require some 
means to deal with the flash of low altitude laser light that follows immediately after each 
laser pulse.  These systems also need a fast shutter to isolate the high altitude portion of 
the focused laser beam to make it appear star-like to the wavefront sensor.  We describe 
how these tasks are accomplished with UnISIS, the Rayleigh laser guided adaptive optics 
system at the Mt. Wilson Observatory 2.5-m telescope.  We use several methods:  a 
10,000 RPM rotating disk, dichroics, a fast sweep and clear mode of the CCD readout 
electronics on a 10 $\mu$s timescale, and a Pockel's cell shutter system.  The Pockel's cell 
shutter would be conventional in design if the laser light were naturally polarized, but the 
UnISIS 351 nm laser is unpolarized.  So we have designed and put into operation a dual 
Pockel's cell shutter in a unique bow tie arrangement.

\end{abstract}


\section{INTRODUCTION}

The UnISIS Rayleigh laser guide star system has been commissioned as described in 
\citet{tho02}.  The system is built around a 30 Watt excimer laser that emits short 90 mJ 
pulses of 351 nm light at rates up to 333 Hz.  These pulses are projected in the "full-
aperture broadcast" mode and are focused in the stratosphere at an altitude of $\sim$20 
km above mean sea level ( $\sim$18 km above the telescope).  Immediately after the 
outgoing pulse of laser light hits the telescope primary mirror, a bright flash of low 
altitude Rayleigh scattered light fills the near-field of the telescope.  As described below, 
the science cameras can be shielded from this near-field burst of light by dichroics 
because the projection method -- a 10,000 RPM rotating glass disk -- blocks the adaptive 
optics system from seeing a large fraction of this light in the first few microseconds after 
the laser pulse.  It is more difficult to protect the laser guide star wavefront camera 
because it is highly sensitive to the laser wavelength.  We use two methods to accomplish 
the latter task:  a continuous high speed read and flush of the CCD wavefront camera and 
a Pockel's cell shutter.
 
We start in Sec. 2 with a general discussion of ways that other laser guide star projection 
systems handle the contaminating Rayleigh scattered light, and then we review the 
UnISIS rotating disk projection system.  In Sec. 3 we describe briefly the dichroic 
isolation of the UnISIS science cameras.  In Sec. 4 we describe how the UnISIS 
wavefront sensor is hidden from the low altitude near-field burst of Rayleigh light with 
our new bow tie Pockel's cell shutter system, and then in Sec. 5 we describe the second 
level of protection for the wavefront sensor:  a continuous high-speed read and flush 
mode of the UnISIS wavefront CCD.  The general characteristics of the UnISIS CCD 
system and its future upgrades are also described in Sec. 5. An up to date block diagram 
of the UnISIS optical layout will be included in the next paper of this series, but a 
close approximation (not showing the bowtie shutter) can be found in \citet{tho98}.

\section{LASER GUIDE STAR PROJECTION SYSTEMS}

Sodium resonance line excitation at 589.3 nm is the primary alternative to Rayleigh laser 
guide stars, yet sodium laser guide star systems also suffer from low altitude Rayleigh 
contamination.  Because astronomy-qualified sodium laser systems that have been built 
to date are continuous lasers (or at least quasi-CW), design alternatives are limited for 
hiding their low altitude contaminating Rayleigh scattered light.  A commonly suggested
scheme for sodium lasers is to project the laser beam from behind the 
Cassegrain secondary mirror with special purpose projection optics.  Once this scheme is
implemented, the Cassegrain secondary mirror will provide a natural shadow when low 
altitude Rayleigh light is viewed from the 
telescope focal plane.  As laser guide star systems become more complex, with multiple 
laser beams projected from the same telescope, only the lowest parts of the Rayleigh 
scattered light can be hidden behind the Cassegrain mirror.  The two simple (single laser 
guide star) sodium systems now in operation or close to it -- at Lick and Keck Observatories 
-- employ quasi-CW lasers, and the laser projector is a small collimator strapped to the 
side of each telescope.  In these two cases, Rayleigh scattered light is visible in the 
telescope focal plane, but it appears projected off to the side of the science object.  A 
focal plane mask blocks it out.  With future multiple laser guide star systems with 1 
arcmin to 2 arcmin fields of view, focal plane masking will not work, so the low altitude 
Rayleigh scattered light is likely to become an obstacle to the installation of multiple 
laser guide star systems based on CW sodium-wavelength lasers.

The advantages of the full aperture broadcast method of UnISIS were described by 
\citet{tho02} who termed the UnISIS system "Stealth qualified" because the FAA has 
placed no restrictions on the use of the laser projection system, and they require no active 
countermeasures.  At Mt. Wilson a major fraction of the 2.5-m primary mirror is used as 
the projector optic, and this feature of UnISIS helps to dilute the energy density at all 
points except at high altitude where the laser is focused.  Given the expected size of the 
Cassegrain mirror on the next generation of Extremely Large Telescopes, laser projection 
optics nearly as large as the 2.5-m primary mirror at Mt. Wilson could be used.  This 
would allow these systems to be Stealth qualified, too, and would hide the lowest 
portions of the outgoing Rayleigh scattered light in the shadow of the central Cassegrain 
obstruction.  But for the present, the UnISIS "Stealth" characteristic comes at the price of 
having Rayleigh scattered light projected onto the primary mirror of the telescope in full 
view of the science cameras.

The Rayleigh guide star system at Starfire Optical Range \citep{fug94} also used the 
telescope primary mirror to project a pulsed laser beam into the sky in full aperture 
broadcast mode.  The polarized Starfire laser beam was injected into the telescope by 
placing a polarizing beamsplitter cube on the telescope's optical axis.  Light from a 
pulsed copper-vapor laser was directed into the side of this cube, and light from satellites 
and astronomical objects was viewed through it in a direct line along the telescope's 
optical axis.  Not surprisingly, the cube beamsplitter fluoresced continuously, and the 
background light it emitted prevented the Starfire system from seeing low surface 
brightness astronomical objects.

To avoid the fluorescence encountered in the Starfire system, UnISIS is designed to have 
as its beam-sharing element a 10,000-RPM rotating disk.   Highly reflective and laser 
hardened multi-layer dielectric spots are deposited onto the front surface of a glass disk, 
and the laser light is reflected off these spots.  In principle, no light enters the glass disk 
substrate.  Two other effects are also at work.  First, even if the substrate beneath the 
reflective spots glows by fluorescence (excited by photons that leak through the reflective 
spot), the fluorescing portion of the disk rotates off the optical axis.  Second, for the short 
time that the spot remains on the optical axis (as the laser fires), it acts like a shutter to 
prevent near-field Rayleigh scattered light from entering the UnISIS adaptive optics 
system.  The laser pulse length is $\sim$20 ns and the disk rotates completely out of the 
beam over a period of $\sim$100 $\mu$s.  For the first few microseconds (50 times 
longer than the laser pulse and at a time when the Rayleigh backscatter is the strongest) only a 
small fraction of the pupil opens as the reflective spot moves to the side.  This shuttering 
action of the reflective spot helps to hold back the 351 nm photons for both the dichroic 
shield erected for protection of the science cameras and the electronic shutter used for the 
wavefront sensor CCD.  For more details on the rotating disk design and its operation, 
see \citet{tho95} and \citet{tho02}.

\section{HIDING THE SCIENCE CAMERAS FROM RAYLEIGH SCATTERED 
LIGHT}

The first several optical elements in the UnISIS adaptive optics system are shared for 
both the science and the down-link UV laser guide star photons.  The dichroic separation 
is made immediately after the UnISIS deformable mirror.  At this point, the UV light is 
reflected off an optical surface coated with a multi-layer dielectric with peak reflectivity 
greater than 99.9\% at 351 nm, and the longer wavelength science light passes through it.  
This special-order dichroic (Optical Coating Technologies, Inc., Easthampton, MA) is 
flat topped and  $\sim$20 nm wide.  Wavelengths longer than 400 nm are transmitted to 
the science focal plane, and we count on UV blocking filters (for the CCD science 
camera) and the lack of UV and blue sensor responsivity (for the near-IR camera) to 
suppress all laser guide star photons that manage to leak to the final science focal plane of 
UnISIS.

\section{POCKEL'S CELL SWITCH AND THE BOW TIE}

Much more care must be taken to prevent laser contamination in the UV wavefront 
section of UnISIS because there the optics are designed to have the highest possible UV 
throughput at 351 nm.  As mentioned above, a two-stage plan was devised to block the 
low altitude Rayleigh scattered contaminating light.  The first stage involves a Pockel's 
cell shutter system designed by \citet{cra02}.  In brief, a Pockel's cell shutter consists of 
two crossed polarizers with a Pockel's cell switch sandwiched between them.  When the 
Pockel's cell is turned off, it has no effect on the light, and the crossed polarizers pass no 
light.  When a precisely adjusted level of high voltage is applied to the Pockel's cell's 
birefringent crystal, all light entering the cell has its plane of polarization rotated 90$^o$, 
so 100\% of the light entering the first polarizer exits the second.  The Pockel's cell 
can be switched on and off in microseconds.  A crucial part of a Pockel's cell shutter 
design is to create a precisely collimated beam.  \citet{cra02} reports that the level of 
collimation has to be within one part in 10$^5$ to produce good extinction.

In its first implementation, the Pockel's cell switch used only 50\% of the laser guide star 
return signal.  The UnISIS excimer laser is unpolarized and a conventional Pockel's cell 
switch requires a polarized input beam.  When the UnISIS return laser guide star flux was 
first measured, this flux was too low to operate the wavefront sensor in closed loop.  At 
that point it was obvious that some means would have to be devised to recover the lost 
50\% of the laser guide star signal.  The bow tie design allows both polarizations to enter 
the wavefront camera.  

The bow tie is based on the birefringent properties of Thompson polarizers.  Figure 1 
shows the functional properties of a single Thompson polarizer which consists of two 
pieces of calcite cemented together and cut in such a way that one entering unpolarized 
beam produces two output beams, one of each polarization.  These two beams emerge 
with an angular separation of $\sim$45$^o$.  Figure 2 shows a schematic drawing of the dual 
Pockel's cell bow tie switch from \citet{cra02}.  Note that all optical components in 
Figure 2 sit in a precisely collimated portion of the UV wavefront optical train.  The first 
Thompson prism, positioned at the entrance to the bow tie, works in a direct way to split 
the single entrance beam into two orthogonally polarized beams.  The second Thompson 
prism, positioned at the exit of the bow tie, performs the reverse function of recombining 
the two polarized beams.  The two separate paths through the bow tie satisfy the Pockel's 
cell design requirement of having a Pockel's cell sandwiched between two crossed 
polarizing elements.  

In Figure 2, the bow tie appears to be symmetric with the two beams exiting the first 
Thompson prism at exactly 45$^o$.  In reality, the angle is not exactly 45$^o$ because 
commercially available Thompson prisms are designed to work over a relatively broad 
wavelength range, and the calcite (like all optical materials) does not have the same index 
of refraction at all wavelengths.  The calcite crystals are cut to make the angle between 
the two output beams close to 45$^o$.   Figure 3 shows the bow tie Pockel's cell switch as it 
is implemented on the UnISIS optics table.  Note the 180$^o$ reversal (top to bottom) 
between the design drawing and the implementation:  the exit beam line is along the top 
of Figure 2 but along the bottom of Figure 3.  The two Pockel's cells were manufactured 
by Inrad Incorporated (Northvale, NJ) and are Model 212-150 with KD*P crystals.  They 
are switched with a TTL pulse sent to a Q-switch driver, which is also from Inrad, Model 
2-018 with high voltage adjustable up to 5 kV.  The 1/4 wave phase delay occurs at 
approximately 3.5 kV with both Pockel's cells. 

The installation and alignment of the bow tie optical system required tremendous 
patience.  The two output beams must be recombined with the two optical axes 
coincident and the beam directions identical.  The pupil images must coincide to 
within a fraction of the 24$\mu$m pixel size on the wavefront sensor.  Attaining this 
adjustment is eased somewhat by a beam reduction between the 5 mm diameter 
collimated beam and the 0.936 mm pupil diameter on the wavefront CCD.  The two 
collimated beams must match to $\sim$20 $\mu$m.  Once this alignment is achieved on 
the UnISIS optics table, it seems to hold for months.  

Maximum extinction is reached when a Pockel's cell is oriented exactly perpendicular 
to the local optical axis. A gross misalignment destroys the extinction, and a minor misalignment 
produces irregular pupil illumination. As installed in UnISIS, the bow tie produces a maximum 
extinction of 1000.

At the present time, the bow tie shutter has one deficiency: its front-to-back transmission
is only 34\% (i.e. a single leg of the bow tie transmits 17\%). We plan to investigate three 
areas that might lead to improved throughput. First, each of the 
Thompson prisms consists of two pieces of calcite held together with cement. We suspect that 
the cement within the Thompson prisms might have unusually high absorbtion at 351 nm. Calcite 
is a natural product, and heavy element contaminants in optical materials absorb UV photons. Third,
the 351 nm absorbtion might be high simply because the optical path length through the calcite is 
high. In its current configuration, the bow tie shutter requires each beam to traverse 2 x 30 mm of
calcite.

In an alternate design for the bowtie shutter, the Thompson prisms could be replaced with Glan-Laser 
prisms. These prisms also split an unpolarized input beam into two polarized components and separate
them by approximately 45$^o$. Although Glan-Laser prisms are also made of calcite, the path 
length for each beam can be as small as 2 x 13 mm, and the two pieces of calcite are not cemented: they
are separated by an air-space. A pair of Glan-Laser prisms placed at opposite corners of a parallelogram
will mimic the function of the bowtie switch described here.

We recently tested the transmittance of a 10 mm square Glan-Laser prism provided by Nova Phase, Inc. 
(Newton, NJ). After accounting for the loss of light at the (currently uncoated) prism surfaces, the 
internal transmittance for this particular calcite prism was found to be 30$\%$ for a single polarization. This 
should be compared to a 17$\%$ transmittance for one leg of the current system with Thompson prisms. Finally,
we note that a completely separate alternate strategy for shuttering the pulsed Rayleigh return signal is
discussed at the end of Section 6.

\section{WAVEFRONT SENSOR ELECTRONICS AND THE CCD39}

The Marconi CCD39 is well suited to wavefront sensing applications. It has a relatively small 
format at 80x80 pixels, 4 read-out amplifiers, relatively high UV quantum efficiency, and 
low amplifier read noise.  The light sensitive area is 1.92x1.92 mm (see Figure 4).  
Because the flash of low altitude Rayleigh light is very bright, a significant level of 
contaminating light leaks through the Pockel's cell shutter.  To insure that the 
contaminating signal is gone by the time the return laser guide star light arrives $\sim$100$\mu$s 
later, the following clearing process was implemented in the camera electronics (built by 
Astronomical Research Cameras Incorporated, San Diego, CA).  When the CCD39 is 
not collecting an exposure, it is programmed to idle in a fast and continuous readout 
mode in which both parallel and serial registers on all four amplifiers are cleared.  In this idle 
period, these registers are read-out at the full 10 MHz clock rate, and the output 
information is simply dumped.  This process continues until the camera receives an 
external TTL signal, at which point the sensor is immediately reset and put into an 
integration mode.  

As explained in \citet{tho02}, a closed-loop wavefront correction cycle of UnISIS 
begins when a TTL pulse is created as an optical fiber detects the outgoing laser pulse.  
The TTL pulse is sent into a digital delay generator programmed to wait 100$\mu$s before 
strobing the CCD39 which halts the CCD clearing mode and prepares the sensor to 
receive the laser guide star light.  After the camera settles for about 10$\mu$s, the Pockel's 
cell shutter is opened for the time gate appropriate for the laser guide star return signal.  
The CCD39 then finishes its exposure, the camera readout begins, and this readout 
triggers the adaptive optics reconstructor computer to analyze the wavefront gradients 
from the Shack-Hartmann image it has just received.

Experimental tests of the CCD39 sensor with the Pockel's cell shutter show that both 
systems (CCD clearing and Pockel's cell extinction) are needed to hold back the 
Rayleigh backscatter flash.  An example of the wavefront return signal from the Rayleigh 
laser guide star is displayed in Fig. 10 of \citet{tho02} where we display a single wavefront
snapshot as well as the integrated sum of 45 consecutive exposures 
from a 20 Hz duty-cycle experiment.  In this experiment, the original UnISIS CCD39 
sensor was being used with its 48\% quantum efficiency and a read noise of 4.4e$^-$ 
rms at a 3.2$\mu$s per pixel read rate.  A new sensor is now available with 68\% quantum 
efficiency, and because the new sensor has a somewhat higher amplifier responsivity, we 
anticipate the same or perhaps lower read noise with the new sensor.  Mt. Wilson is a site 
where RF interference is a problem, so it was a major accomplishment to get the 
CCD39 to perform at the 4.4e$^-$ r.m.s. noise levels.  This is the same performance we 
see in an RF-free lab setting.  By trial and error we found that the lowest noise was 
achieved when the camera and electronics were completely isolated from all ground 
planes.

In its current configuration, the UnISIS wavefront sensor works at a maximum frame rate 
of 833 frames per second giving a frame to frame latency of 1.2ms in closed-loop 
operation.  Note that the per pixel dwell time of 3.2$\mu$s mentioned above applies only 
to the 13x13 array of active Shack-Hartmann quadcells (each 2x2 pixels), so the actual 
read time for the active pixels is 0.54$\mu$s.  The camera electronics consume the 
remaining 0.7$\mu$s.  New timing board electronics now available through 
Astronomical Research Cameras Incorporated will allow the UnISIS CCD camera frame 
rate to increase to 1360 frames per second for a frame-to-frame latency of 735$\mu$s.  
This upgrade will significantly improve the closed-loop performance of UnISIS on nights 
when the atmospheric turbulence has a short timescale.

The 13x13 quadcells used by UnISIS are distributed over an area on the sensor that spans 
38x38 pixels because guard rows and guard columns separate each active 2x2 quadcell to 
prevent spillover.  During a read-out cycle of the exposed frame, the guard columns and 
rows are skipped and the data are dumped at a high rate much like the clearing mode 
described above.  A map of the active pixels superimposed on the 80x80 sensor geometry 
is shown in Figure 5 (reproduced from \citet{tho98}).

\section{SUMMARY}

Laser guided adaptive optics is still in an early phase of development, and there are many 
ways yet to be found to configure both the lasers and the cameras to minimize 
interference from the inevitable low-altitude Rayleigh scattered light, whether this comes 
from sodium resonance laser guide stars or from Rayleigh laser guide stars.  We have 
presented here several examples of how this is accomplished in UnISIS.  We find that the 
greatest design flexibility comes with pulsed lasers rather than CW systems, and one of 
the more significant characteristics of the system design is whether aircraft and satellite 
avoidance can be handled in a fashion that can be called "Stealth".  Pulsed lasers at 351 
nm provide a distinct advantage in both regards.  However, new technological 
innovations will continue to alter the balance for some time to come.  One simple 
example is the potential ability of embedding the shutter function in the silicon substrate of 
the wavefront sensor CCD, thereby making Pockel's cell switching unnecessary.  MIT / 
Lincoln Laboratory has the ability to produce sensors of this type, but no UV sensitive 
versions are currently available for wavefront sensing \citep{rei02}.  Other technological advances 
will, no doubt, come along as laser guided adaptive optics systems mature.

A number of people have contributed to the work reported here.  This includes Richard 
Castle, Dr. E. Harvey Richardson and Bill Knight and his machine shop crew.  At Mt. 
Wilson Observatory we acknowledge support provided by Mount Wilson Institute Director 
Dr. Robert Jastrow and technical assistance by Robert Cadman, Sean Hoss, Chris Hodge, 
Joe Russell, Victor Castillo and Thomas Schneider of Schneider Engineering.  Telescope 
operator support at the 2.5-m telescope was provided by Kirk Palmer, Michael Bradford, 
and Jim Strogen.  Marconi CCD39 wavefront software support was provided by Jamie Erickson and 
Scott Striet.  The Pockel's cells used in UnISIS were supplied by ThermoTrex 
Incorporated (San Diego, CA) as part of a surplus equipment transfer, and we thank Dr. 
David Sandler for assistance in that transfer.  This work was supported by grants from the 
National Science Foundation:  AST-9220504, AST-0096741 and by funds from both the 
University of Illinois and the New Mexico Institute of Mining and Technology.  All 
support is very gratefully acknowledged.

\clearpage
\begin{figure*}
\epsscale{0.5}
\plotone{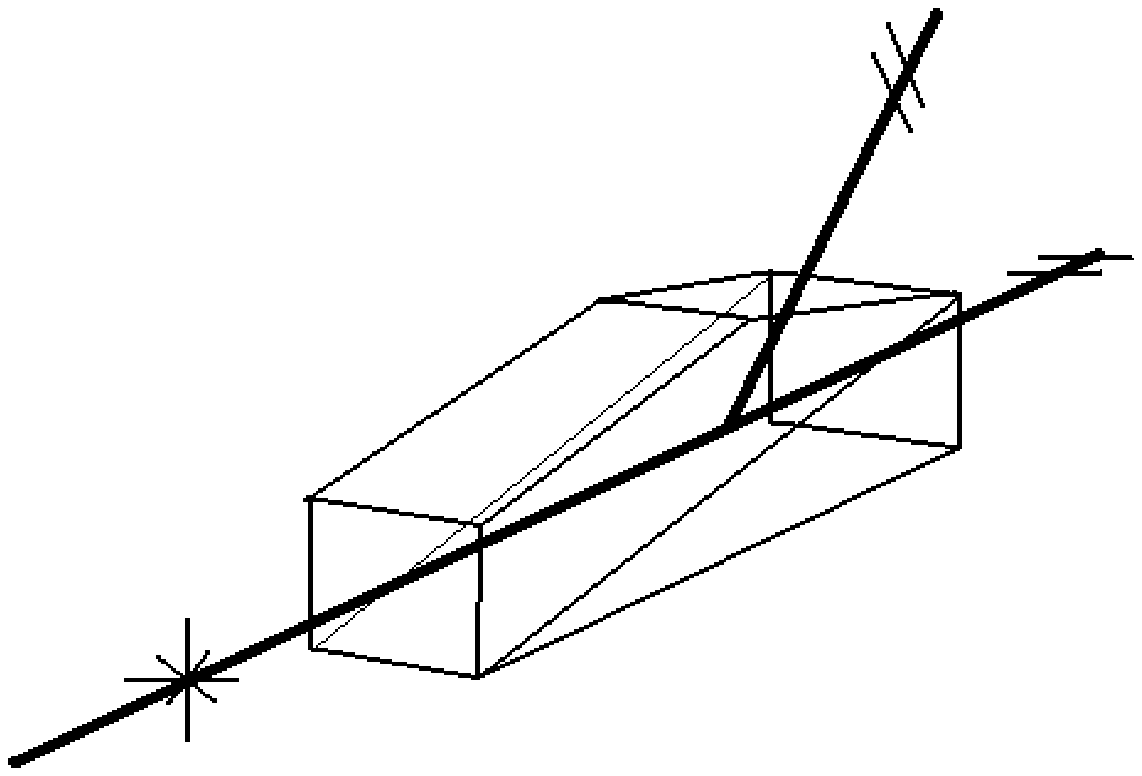}
\caption{The functional characteristics of a Thompson prism can be seen in this 
schematic drawing.  Unpolarized light enters normal to the left hand face, and one 
component exits the far side without changing direction.  The other polarized component 
is bent upwards in this diagram and exits the top face at an angle $\sim$45$^o$ from the input 
beam.  The Thompson prism consists of two pieces of calcite cemented at their interface.  
The bottom piece has a triangular cross-section and the top piece a trapezoidal cross-
section.  In the bow tie Pockel's cell switch, all beams entering and exiting the two 
Thompson prisms are parallel to the optics table.
\label{fig1}}
\end{figure*}

\clearpage
\begin{figure*}
\epsscale{0.5}
\plotone{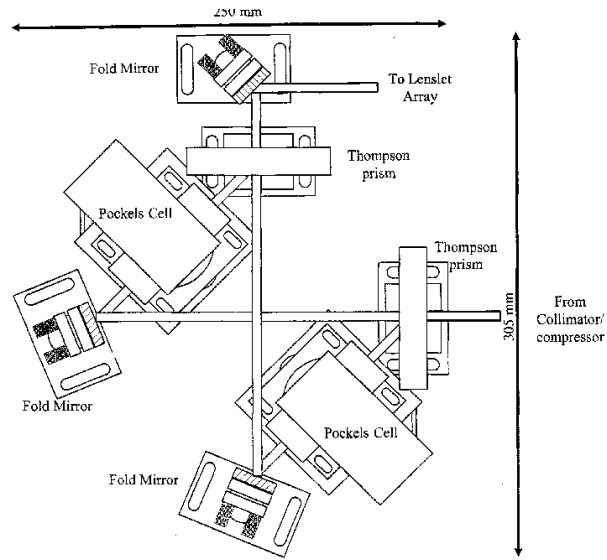}
\caption{Schematic drawing of the bow tie Pockel's cell switch from Crawford (2002).  
The entrance beam from the sky comes from the right hand side, and the exit beam to the 
wavefront CCD camera exits the diagram at the position labeled "To Lenslet Array".
\label{fig2}}
\end{figure*}

\clearpage
\begin{figure*}
\epsscale{0.5}
\plotone{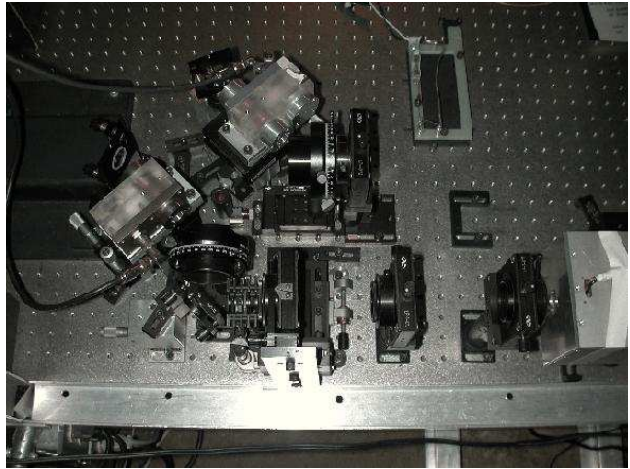}
\caption{Bow tie Pockel's cell switch as implemented on the main UnISIS optics table.  
Be aware that the design (from Figure 2) was flipped top-to-bottom when the components 
were installed on UnISIS.  In this picture the beam from the sky enters the diagram from 
the right.  The Thompson prisms are held in the two circular black mounts with 360 
angular scales around their perimeter.  The two Pockel's cell crystals are housed inside 
clear plastic cases (to hold back the high voltage) and are each fed with a HV cable.  The 
exit beam to the wavefront CCD runs through the lens holders along the bottom of the 
diagram towards the right.  The CCD39 sensor is mounted inside the cubical 
aluminum box in the lower right corner of this picture.
\label{fig3}}
\end{figure*}

\clearpage
\begin{figure*}
\epsscale{0.5}
\plotone{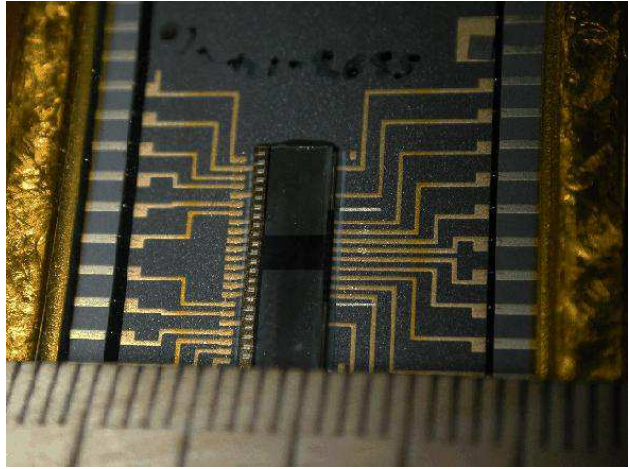}
\caption{Front surface of the CCD39 is shown in this image.  The light sensitive 
area sits in the middle of the dark band at the center of the picture.  With 24$\mu$m pixels 
and an 80x80 format, the sensor is a 1.92x1.92 mm square.  Note the slightly out of focus 
mm scale in the foreground.
\label{fig4}}
\end{figure*}

\clearpage
\begin{figure*}
\epsscale{0.5}
\plotone{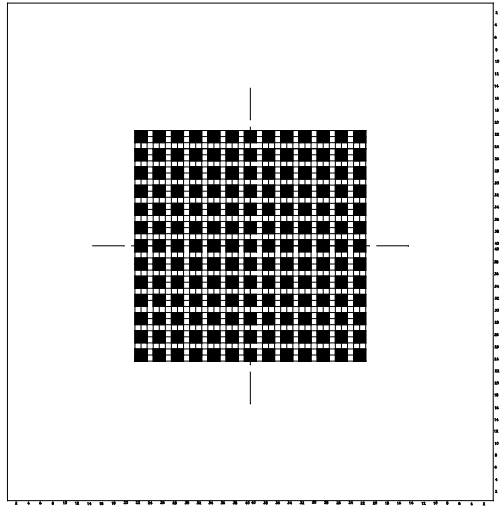}
\caption{The outer square border represents the full extent of the 80x80 pixels in the 
CCD39 sensor.  Each small black square represents a 2x2 pixel quadcell.  The 13x13 
array of black boxes represents the array of Shack-Hartmann subimages used in the 
operation of UnISIS.  When the signal from any black area reaches an amplifier, the 
camera electronics spends 3.2$\mu$s reading that pixel.  When a the "signal" from a white 
area reaches the amplifier, it is dumped as quickly as the electronics will allow.  The 
white areas between quadcells are "guard rings" to prevent stray signals from adjacent 
Shack-Hartmann subimages from accidentally reaching adjacent quadcells.  The CCD39
sensor has four amplifiers, one for each quadrant of the detector.  Short lines in the 
central part of the diagram designate sensor quadrant boundaries.
\label{fig5}}
\end{figure*}

\end{document}